\documentstyle[multicol,aps,prl,epsf,psfig]{revtex}

\begin{document}
\title{Multiplicity of Ordered Phases in Frustrated Systems\\
Obtained from Hard-Spin Mean-Field Theory}
\author{H\"useyin Kaya $^{1,2}$ and A. Nihat Berker $^{1,2,3}$}
\address{$^1$ Feza G\"ursey Research Institute for Basic Sciences, 81220 \c Cengelk\"oy,
Istanbul, Turkey}
\address{$^2$ Department of Physics, Istanbul Technical University, 80626 Maslak,
Istanbul, Turkey }
\address{$^3$ Department of Physics, Massachusetts Institute of Technology, Cambridge,
Massachusetts 02139, USA}
\maketitle

\begin{abstract}
Random quenched dilution of the triangular-lattice antiferromagnetic Ising
model locally relieves frustration, leading to ordering phenomena. We have
studied this system, under such dilution of one sublattice, using hard-spin
mean-field theory. After a threshold dilution, two sublattices develop
non-zero magnetizations of equal magnitude and opposite signs, as all three
sublattices exhibit spin-glass order. In this phase, multiple sets of
ordered solutions occur. A phase diagram is obtained in dilution fraction
and temperature.

{\underline {PACS Numbers}:} 75.10.Nr, 05.70.Fh, 64.70.Pf, 75.30.Cr

\end{abstract}

\begin{multicols}{2}

A ''rugged free-energy landscape'' is often mentioned as a distinctive
characteristic in the discussions of spin-glass systems~\cite{rugged}. 
Concrete support for such a phenomenon would derive from multiple
solutions, not related to each other by a global symmetry, of
self-consistent order-parameter equations. This has not been previously
obtained for any system with a realistic spatial connectivity. In this
work, we do find such multiple solutions, not related by global
symmetry, in a random frustrated system with realistic spatial
connectivity, namely the quench-diluted triangular-lattice
antiferromagnetic Ising model, studied via the closed-form
implementation of hard-spin mean-field theory\cite
{Netz1,Netz2,Banavar,Netz3,Netz4,Netz5,Berker,Alkan,Akguc,Monroe,Alkan2}. 

The antiferromagnetic Ising model, with Hamiltonian 
\begin{equation}
\label{hamil}-\beta {\cal H}=-J\sum_{<ij>}s_is_j,\,\,\,\,\,\,\,\,\,J\ge 0, 
\end{equation}
where $s_i=\pm 1$ at each site $i$ of a triangular lattice and $<ij>$
indicates summation over nearest-neighbor pairs of sites, is fully
frustrated~\cite{Toulouse}: In each elementary triangle, one of the
three nearest-neighbor antiferromagnetic interactions is dissatisfied
when the energy is minimized. This leads, macroscopically, to a highly
degenerate system that is disordered at all non-zero $(1/J>0)$
temperatures\cite {Wannier}. Random quenched dilution of the system
relieves the frustration at random localities and can be expected to
lead to ordering phenomena.  Indeed, a Monte Carlo study with random
quenched dilution of all three sublattices equivalently has indicated
spin-glass order~\cite{Grest}. We consider the random quenched dilution
of the sites of one of three sublattices. After a threshold dilution,
the system exhibits uniform and opposite magnetizations in two
sublattices and spin-glass order, i.e., spins frozen in random
directions, in the quench-diluted sublattice. A phase diagram is
obtained in the variables of dilution fraction and temperature.  Within
the ordered phase, for a fixed dilution fraction and temperature, a
multiplicity of solutions, distinguished by different values of the
local and global order parameters, is obtained to the hard-spin
mean-field equations. 

We use hard-spin mean-field theory, a method which is almost as simply
enunciated as usual mean-field theory, but which conserves
frustration~\cite
{Netz1,Netz2,Banavar,Netz3,Netz4,Netz5,Berker,Alkan,Akguc,Monroe,Alkan2}. 
Consequently, this method has yielded, for example, the lack of order in
the undiluted triangular-lattice antiferromagnetic Ising model and the
ordering that occurs when a uniform magnetic field is applied to the
system, in a quantitatively accurate phase diagram in the temperature
versus magnetic field variables
\cite{Netz1,Netz2,Banavar,Netz3,Berker,Alkan}. Hard-spin mean-field
theory also yields the lack of order in the one-dimensional Ising
ferromagnet and the occurrence of order in the two-dimensional Ising
ferromagnet, the latter with an onset temperature improved over usual
mean-field theory \cite{Banavar}. Hard-spin mean-field theory has also
been successfully applied to complicated systems that exhibit a variety
of ordering behaviors, such as three-dimensional stacked frustrated
systems \cite {Netz1,Netz4} and higher-spin systems \cite{Netz5}. 

The self-consistent equation for local magnetizations in hard-spin
mean-field theory is 
\begin{equation}
\label{hspin}m_i=\sum_{\{s\}}\biggr[\prod_jp(m_j;s_j)\biggl]\tanh \bigl( 
-J\sum_{j^{\prime }}s_{j^{\prime }}\bigr),
\end{equation}
where the product over $j$ and sum over $j^{\prime }$ run over all
non-diluted sites neighboring site $i$, and the single-site spin
probability distribution $p(m_j;s_j)$ is $(1+m_js_j)/2$. The outer
summation is over the $\pm 1$ values of the spins at the undiluted sites
neighboring site $i$.  Thus, the spin at each site feels the
anti-aligning field due to the full (i.e., hard-) spin of each of its
neighbors. Eq.({\ref{hspin}}) is a set of coupled equations for all the
local magnetizations and is solved iteratively for a given realization
of dilution in a finite but large system. 

Alternatively, {\it a further approximation} is to impose sublatticewise
uniformity, $m_j=m_\alpha $ for each sublattice $\alpha =a,b,c$, and to
average the self-consistent equation over all realizations of quenched
site dilution,

\begin{equation}
\label{hspinav}m_i=\sum_{\{\eta \}}Q(\{\eta \})\left\{ \sum_{\{s\}}\biggr[
\prod_jp(m_j;s_j)\biggl]\tanh \bigl( -J\sum_{j^{\prime }}s_{j^{\prime }}
\bigr)\right\} .
\end{equation}
In Eq.({\ref{hspinav}}), the parentheses enclose the right-hand side of
Eq.({\ref{hspin}). This is summed over the }${2}^6$ possible quenched
environments $\{\eta \}$ of site $i$. Each quenched environment $\{\eta
\}$ occurs with a probability $Q$ composed of six factors $q_j$, with $
q_j=(1-p_j)$ for each quench-diluted neighbor $j$ and $q_j=p_j$ for each
undiluted neighbor $j$.  Eq.({\ref{hspinav}}) can be compactly rewritten
as

\begin{equation}
\label{hspinav2}m_i=\biggl[\prod_j\sum_{\eta _j=0,1}q(p_j;\eta
_j)\sum_{s_j=\pm 1}p(m_j;s_j)\biggr]\tanh (-J\sum_{j^{\prime }}\eta
_{j^{\prime }}s_{j^{\prime }}),
\end{equation}
where the product over $j$ and sum over $j^{\prime }$ run over all sites
neighboring site $i,$ the single-site quenched-dilution probability
distribution $q(p_j;\eta _j)$ is $1-p_j-\eta _j(1-2p_j)$, and
$p_j=p_\alpha $ for each sublattice $\alpha =a,b,c$.
Eq.({\ref{hspinav}}) or, equivalently, Eq.({\ref{hspinav2}}) is solved
for $(m_a,m_b,m_c)$ for given $(p_a,p_b,p_c)$ . 

Whereas hard-spin mean-field theory [Eq.(\ref{hspin})] yields the
variations in the local magnetizations within each sublattice due to
differently quenched local environments, the implementation of
Eq.({\ref{hspinav}}) is a further approximation over hard-spin
mean-field theory: While still incorporating frustration, it imposes
sublatticewise uniform magnetizations.  Eq.({\ref{hspinav}}) is a set of
three coupled equations, whereas Eq.({\ref {hspin}}) is a set of $N$
coupled equations, where $N$ is the size of the system. 

Upon random quenched dilution of one sublattice, the frustrated
triangular-lattice Ising model does indeed show long-range order, as for
example depicted in Figs.1. The two undiluted sublattices (labeled $a$
and $b$), which are now subject to random unfrustrated localities at
the dilution points of the other sublattice (labeled $c$), develop
non-zero sublattice-averaged magnetizations, $m_a=-m_b$, at low
temperatures. For low dilutions, these magnetizations [Fig.1(a)] show an
initial slow growth at onset as temperature is lowered and do not
saturate at zero temperature. The diluted sublattice $c$ is also subject
to local liftings of frustration, due to the spatially non-uniform
magnetizations of sublattices $a$ and $b$, but this is a secondary and,
therefore, weaker effect, and sublattice $c$ does not develop a non-zero
sublattice-averaged magnetization, $m_c=0$. 

All three sublattices develop, within the ordered phase, non-zero
spin-glass order~\cite{Edwards}, i.e., randomly frozen order, with order
parameters
\begin{equation}
\label{opud}q_\alpha =\biggl[{\frac 1{{N_\alpha }}}\sum_i^\alpha
(m_i-m_\alpha )^2\biggr]^{1/2}, 
\end{equation}
where $N_\alpha $ is the number of spins of sublattice $\alpha $. Note that,
for the quenched-diluted sublattice, 
\begin{equation}
\label{opd}q_c=\biggl[{\frac 1{{N_c}}}\sum_i^cm_i^2\biggr]^{1/2}. 
\end{equation}
At high dilutions, the spin-glass ordering trend shows reentrance (as
temperature is lowered, increases and then decreases) on the undiluted
sublattices [Fig.1(b)] and double reentrance (increase, decrease, and
again increase) on the diluted sublattice [Fig.1(c)]. Maximal
zero-temperature spin-glass order occurs at intermediate dilutions, as
seen in Figs.2(b,c). 

\begin{figure}
\begin{center}
\leavevmode
\psfig{figure=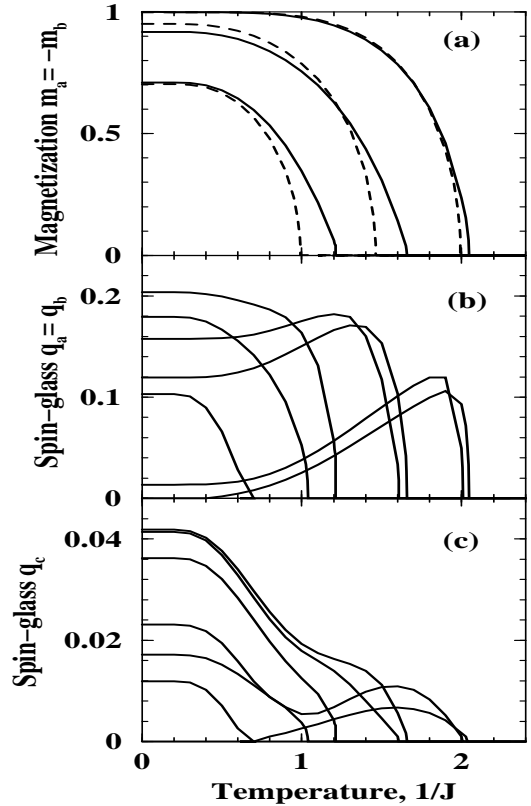,width=7cm,height=11cm}
\end{center}
\narrowtext
\caption{ Finite--temperature order in the random quench\-diluted
triangular--lattice antiferromagnetic Ising model: $(a)$ Magnetizations
of the two undiluted sublattices from hard-spin mean-field theory (full
curves).  The three curves correspond to $p=0.75,0.5,0.09375$, which can
be identified by the respectively increasing critical temperatures,
i.e., increasing x-axis intercepts. The result from the further
approximation of Eq.(3)  is given with the dashed curve. The
quench-diluted sublattice has zero magnetization. $(b)$ Spin-glass order
parameter of the undiluted sublattices.  The curves, again distinguished
by their respectively increasing critical temperatures, are for
$p=0.9375,0.890625,0.75,0.625,0.5,0.140625,0.09375$. Note the reentrant
behavior in the magnitudes, for high dilutions. $(c)$ Spin-glass order
parameter of the quench-diluted sublattice. The values of $p$ are as in
$(b)$. Here, the spin-glass order for $p=0.09375$ is away from zero only
at higher temperatures. The spin-glass order for $p=0.140625$ exhibits
doubly reentrant behavior. The full curves in Fig.1 are for a fixed
realization of the quenched disorder in a $24\times 24$ system.}
\label{fig1} 
\end{figure}

\begin{figure}
\begin{center}
\leavevmode
\psfig{figure=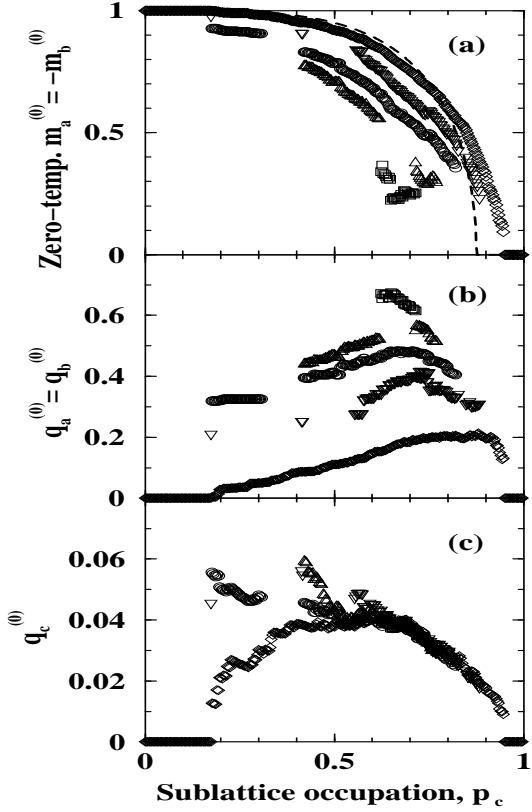,width=7cm,height=11cm}
\end{center}
\narrowtext
\caption{Zero-temperature sublattice magnetizations and spin-glass order
parameters from hard-spin mean-field theory. The sublattice
magnetizations (a) do not saturate at zero-temperature, except for the
full dilution (hexagonal lattice) limit. The dashed curve in (a) is the
result of the further approximation of Eq.(3). The zero-temperature
spin-glass order (b,c) is maximal at intermediate dilutions. Multiple
solutions of the hard-spin mean-field theory equations are seen, in
addition to the most stable set in depicted in Figs.1. (For each
solution, the same symbol is used in Figs.2-6). This figure is obtained
for $1/J=0.0001$, in a $30\times 30$ system.}
\label{fig2}  
\end{figure}

The phase diagram of the system is shown in Fig.3. It is seen that a
threshold dilution of $0.042$ of one sublattice is needed for ordering,
i.e., the occupancy $p$ has to be below $0.958$ for ordering. Fig.3
shows the phase boundaries obtained for two realizations of quenched
dilutions of a $24\times 24$ system and the result from averaging over
$15$ such realizations. Also shown in dashed in Fig.3 is the further
approximation of Eq.(\ref{hspinav}). This dashed phase boundary obeys
the equation 
\begin{eqnarray}
p^3f_3+3p^2(1-&p&)f_2+3p(1-p)^2f_1+(1-p)^3f_0=4/3 \\
{\rm where} \,\,\,\,\,\,\,\,\,f_3 & = & (t_6+6t_4+5t_2)/8,\nonumber \\
f_2 &=&(t_5+3t_3+2t_1)/4, \nonumber\\
f_1 & = &(t_4+2t_2)/2, \nonumber\\
f_0 & = &t_3+t_1,\,\,\,\,\,\,\,\,\,\,\,\,\,\,\,  {\rm where} \,\,
t_n\equiv\tanh
(nJ), \nonumber 
\end{eqnarray}
and gives a dilution threshold of $0.125$.

\begin{figure}
\begin{center}
\leavevmode
\psfig{figure=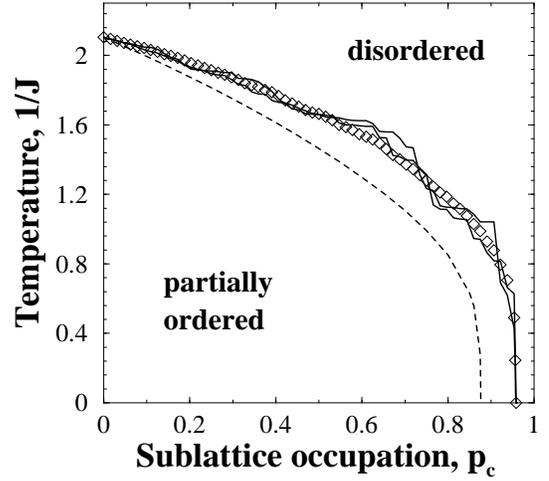,width=7cm,height=6.5cm}
\end{center}
\narrowtext
\caption{Phase diagram of the random quench-diluted triangular-lattice
antiferromagnetic Ising model. The full curves show the phase boundaries
obtained by hard-spin mean-field theory for two realizations of quenched
dilutions of a $24\times 24$ system. The losanges show the result from
averaging over $15$ such realizations. The dashed curve is the result from
the further approximation of Eq.(3); this curve is given analytically by
Eq.(6).}
\label{fig3}
\end{figure}

\begin{figure}
\begin{center}
\leavevmode
\psfig{figure=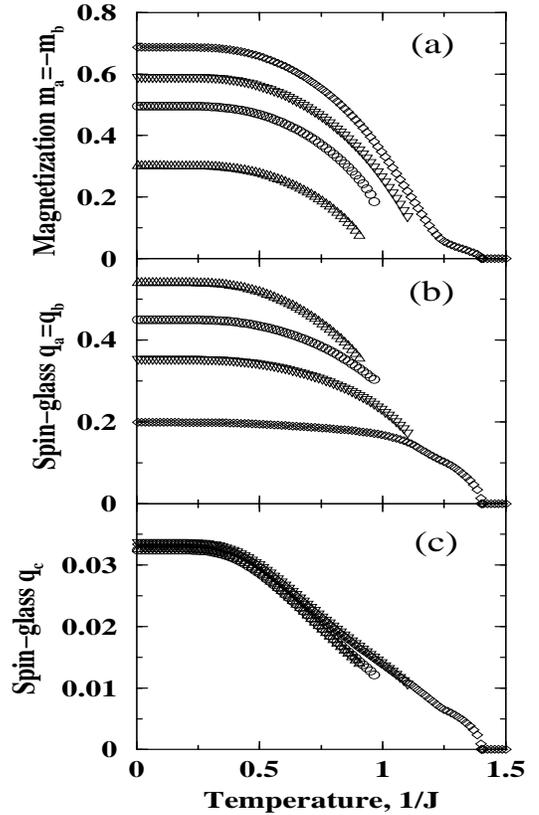,width=7cm,height=11cm}
\end{center}
\narrowtext
\caption{Finite-temperature multiplicity of solutions for $p=0.75$.}
\label{fig4}
\end{figure}

Within the ordered phase, we find that the hard-spin mean-field
equations (\ref{hspin}) admit, as seen in Figs.2,4, a multiplicity of
solutions, in addition to the set depicted in Figs.1. The latter is the
most stable solution, in the sense that it has the largest basin of
attraction under the iterative solution of Eq.(\ref{hspin}). The other
solutions appear at different temperatures below the onset temperature
for the most stable solution. Since, in this study, the number of the
sets of solutions increased in going from the $24\times 24$ system to
the $30\times 30$ system (depicted in Figs.2,4), it can be inferred that
the solutions become numerous in the infinite system.

\begin{figure}
\begin{center}
\leavevmode
\psfig{figure=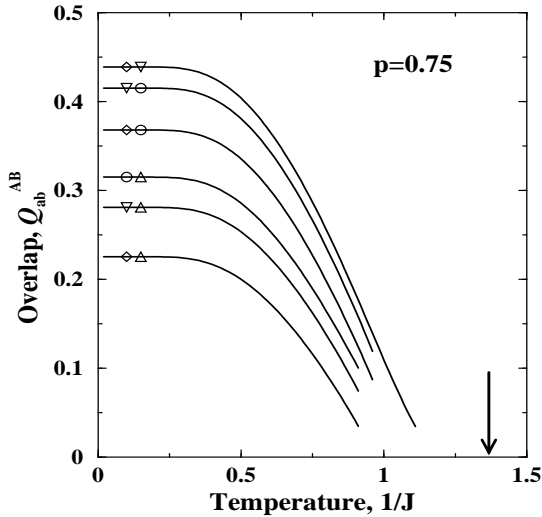,width=7cm,height=7cm}
\end{center}
\narrowtext
\caption{Overlap $Q$ between the different solutions [see Eq.(7a)] on the  
undiluted sublattices, for $p=0.75$. The arrow indicates the
onset of order.}
\label{fig5}
\end{figure}

\begin{figure}
\begin{center}
\leavevmode
\psfig{figure=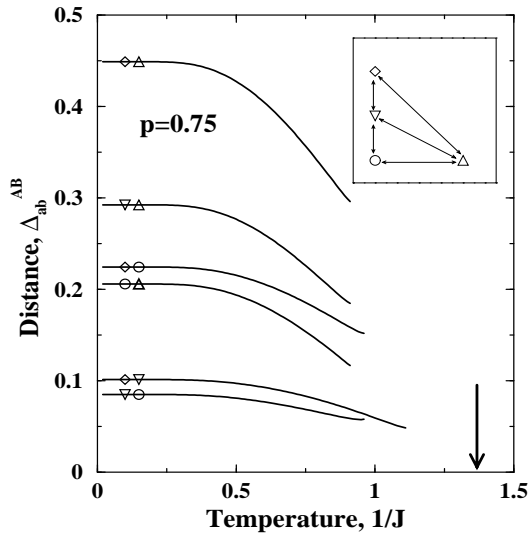,width=6.8cm,height=7.2cm}
\end{center}
\caption{Overlap $\Delta $ between the different solutions [see Eq.(7b)]
on the undiluted sublattices, for $p=0.75$ . The arrow indicates
the onset of order. In inset schematizes the separations between the
different
solutions, showing no ultrametricity.} 
\label{fig6}
\end{figure}

Figs.5,6 depict the overlaps between pairs of solutions $(A,B)$, 
$$
Q_\alpha ^{AB}={\frac 1{{N_\alpha }}}\sum_i^\alpha m_i^Am_i^B \eqno (7a)
$$
and the distances 
$$
\Delta _\alpha ^{AB}={\frac 1{{N_\alpha }}}\sum_i^\alpha (m_i^A-m_i^B)^2. 
\eqno(7b)
$$

From $\Delta _\alpha ^{AB}$ in Fig.6, we can deduce the separation, in
local-order-parameter space, between the different solutions. This is
shown schematically in the inset of the figure. We note that the
different solutions are not ultrametrically\cite{ultrametric} related,
since no isosceles-triangle relations are seen. Thus, it may well be
that ultrametricity is a property specific to the infinitely connected
lattice.

We are grateful to A. Erzan for many valuable discussions. This research
was supported by the Scientific and Technical Research Council of Turkey
(T\"UBITAK) and by the U.S. Department of Energy under Grant No
DE-FG02-92ER45473.

\end{multicols}

\end{document}